\documentclass[twocolumn]{aastex631}

\shorttitle{Variation of magnetic field spectral index in the solar wind}
\shortauthors{McIntyre et al.}

\usepackage{amsmath}
\usepackage{graphicx}

\begin{document}

\title{Properties underlying the variation of the magnetic field spectral index in the inner solar wind} 
\author[0000-0001-9763-9414]{Jack R. McIntyre}

\affiliation{Department of Physics and Astronomy, Queen Mary University of London, London, E1 4NS, UK}

\author[0000-0003-4529-3620]{Christopher H. K. Chen}

\affiliation{Department of Physics and Astronomy, Queen Mary University of London, London, E1 4NS, UK}

\author[0000-0002-7653-9147]{A. Larosa}

\affiliation{Department of Physics and Astronomy, Queen Mary University of London, London, E1 4NS, UK}

\begin{abstract}
Using data from orbits one to eleven of the Parker Solar Probe (PSP) mission, the magnetic field spectral index was measured across a range of heliocentric distances.
The previously observed transition between a value of $-5/3$ far from the Sun and a value of $-3/2$ close to the Sun was recovered, with the transition occurring at around $50 \, R_{\odot}$ and the index saturating at $-3/2$ as the Sun is approached. A statistical analysis was performed to separate the variation of the index on distance from its dependence on other parameters of the solar wind that are plausibly responsible for the transition; including the cross helicity, residual energy, turbulence age and the magnitude of magnetic fluctuations. Of all parameters considered the cross helicity was found to be by far the strongest candidate for the underlying variable responsible.  The velocity spectral index was also measured and found to be consistent with $-3/2$ over the range of values of cross helicity measured. Possible explanations for the behaviour of the indices are discussed, including the theorised different behaviour of imbalanced, compared to balanced, turbulence.

\end{abstract}

\section{Introduction} \label{sec:intro}
The solar wind is known to contain a turbulent cascade and it is proposed that this cascade plays a role in its heating and acceleration \citep{1968ApJ...153..371C, 1971ApJ...168..509B, 1971A&A....13..380A, 2013LRSP...10....2B}. An important diagnostic of the turbulence is the spectral index, $\alpha$, defined by $E(k) \propto k^{\alpha}$, where $E(k)$ is the trace power spectrum and $k$ is wavenumber. The Parker Solar Probe (PSP) mission \citep{2016SSRv..204....7F, 2023SSRv..219....8R} allows us to measure this index across an unprecedented range of heliocentric distances and environments, having already reached a heliocentric distance of less than $14 \, R_{\odot}$. Using data from its first two orbits, \cite{2020ApJS..246...53C} found the magnetic field spectral index, $\alpha_\mathrm{B}$, to vary with heliocentric distance, appearing consistent with $-5/3$ at $0.6 \, \mathrm{au}$ but consistent with $-3/2$ at $0.17 \, \mathrm{au}$. Whether the spectrum would continue to shallow for measurements closer to the Sun was unclear. Using later encounters \cite{2021A&A...650A..21S} and \cite{2023ApJ...943L...8S} found the same transition in $\alpha_\mathrm{B}$ across a wider range of distances.

The two extremes of the transition, $-5/3$ and $-3/2$, are common predictions for $\alpha_\mathrm{B}$ from theoretical models of MHD turbulence. As these predictions are arrived at by making different assumptions about the turbulence, the value of the index can give us insight into its physics. To construct such models it is useful to consider the turbulence in terms of the Elsasser variables, defined as $ \delta \boldsymbol{z}^{\pm} = \delta \boldsymbol{v} \pm \delta \boldsymbol{b}$, where $\delta \boldsymbol{v}$ is the perturbation to the velocity field and $\delta \boldsymbol{b} = \delta \boldsymbol{B} / \sqrt{\mu_0 \rho}$, where $\rho$ is the density, is the perturbation to the magnetic field in velocity units \citep{1950PhRv...79..183E}. From the ideal MHD equations, the evolution of these Elsasser variables is given by
\begin{equation} \label{eq:elasser_evol}
\partial_t \delta \boldsymbol{z}^{\pm} \mp (\mathbf{V}_{\mathrm{A}}\cdot \nabla) \delta \boldsymbol{z}^{\pm} + (\delta \boldsymbol{z}^{\mp} \cdot \nabla) \delta \boldsymbol{z}^{\pm} = -\nabla \tilde{p},
\end{equation} 
where $\mathbf{V}_{\mathrm{A}}$ is the Alfv\'{e}n velocity and $\tilde{p}$ is the total pressure, the sum of the plasma pressure and the magnetic pressure. From this, perturbations to the Elsasser fields can be viewed as wave packets travelling along the background field at the Alfv\'{e}n speed, with $\delta \boldsymbol{z}^{+}$ and $\delta \boldsymbol{z}^{-}$ corresponding to travel in opposite directions. Since the nonlinear term in Equation (\ref{eq:elasser_evol}) requires the presence of both variables to be non-zero, MHD turbulence can be viewed in terms of the interaction of these counter propagating wave packets \citep{1965PhFl....8.1385K}.

In the Iroshnikov-Kraichnan model \citep{1964SvA.....7..566I, 1965PhFl....8.1385K} the turbulence is taken to be isotropic and, in the picture described above, a wave packet must interact with many others to be significantly deformed. In other words the characteristic propagation time of the wave packets is shorter than the nonlinear time of their interactions, this is known as weak turbulence. The resulting spectral index is $-3/2$. However, solar wind turbulence is anisotropic \citep{2012SSRv..172..325H, chen_2016}. This is accounted for in the \cite{1995ApJ...438..763G} model, where the wave packets are elongated along the background magnetic field such that only one interaction of wave packets is necessary to significantly deform those wave packets. Here the turbulence is critically balanced, where the propagation and nonlinear times are taken to be equal --- this condition has been observed to hold in the solar wind \citep{chen_2016}. The result is a $-5/3$ scaling with respect to $k_\perp$, the wavenumber perpendicular to the background field, and a $-2$ scaling with respect to $k_\parallel$, the wavenumber parallel to the background field. Consistent with this, \cite{PhysRevLett.101.175005} reported a $k_\parallel^{-2}$ scaling in the solar wind. To this picture the \cite{2006PhRvL..96k5002B} model adds that the angular alignment of the velocity and magnetic field fluctuations is dependent on scale, giving a $k_\perp^{-3/2}$ scaling and resulting in the wave packets having a three-dimensional anisotropic structure. Observations of the solar wind \citep{2009JGRA..114.1107P, PhysRevLett.110.025003, 2012ApJ...758..120C, 2018ApJ...853...85V} and simulations \citep{2006PhRvL..97y5002M, PhysRevX.2.041005, 2015ApJ...808L..34V, 2016MNRAS.459.2130M} provide mixed evidence for such alignment of fluctuations or 3D anisotropic structure in MHD turbulence.  All these models assume homogeneous background conditions, the picture becomes more complicated when gradients in these conditions are considered \citep{chandran_perez_2019}.  Further, these models assume the energy in the two Elsasser fields to be of comparable magnitude, this is known as balanced turbulence.

It is known that the level of imbalance in energy between the two Elsasser fields varies with heliocentric distance \citep{1987JGR....9211021R, 1995SSRv...73....1T, 1998JGR...103.6521B, 2000JGR...10515959B, 2004GeoRL..3112803M, 2005GeoRL..32.6103B, 2007AnGeo..25.1913B, 2020ApJS..246...53C} and $\alpha_\mathrm{B}$ has been found to depend on the cross helicity, a measure of the imbalance, and residual energy at $1\, \mathrm{au}$ \citep{2010PhPl...17k2905P, 2013ApJ...770..125C, 2013ApJ...778..177W, 2018ApJ...865...45B}. \cite{2023ApJ...943L...8S} found a dependence of $\alpha_\mathrm{B}$ with cross helicity across the distance range provided by PSP. Further, some theoretical models \citep{2007ApJ...655..269L, 2008ApJ...685..646C, 2008ApJ...682.1070B, 2022JPlPh..88e1501S} and simulations \citep{2009ApJ...702..460B, 2010ApJ...722L.110B} suggest imbalanced turbulence behaves differently from balanced turbulence, though this has been disputed \citep{2009PhRvL.102b5003P, 2010PhPl...17e5903P}. The level of imbalance therefore appears as a clear, plausible parameter behind the observed transition in $\alpha_\mathrm{B}$ with distance, however, no previous theoretical work predicts this particular effect. 

In contrast, the velocity spectral index, $\alpha_\mathrm{v}$, has been found to be consistent with $-3/2$ as cross helicity is varied at $1\, \mathrm{au}$ \citep{2010PhPl...17k2905P, 2013ApJ...770..125C, 2018ApJ...865...45B} and to not vary with distance across the distance range provided by PSP \citep{2021A&A...650A..21S}. However, \cite{2010JGRA..11512101R} reported $\alpha_\mathrm{v}$ to evolve with heliocentric distance from $-3/2$ at $1 \, \mathrm{au}$ to $-5/3$ at distances of several au, with some evidence of shallower spectra being associated with regions of high cross helicity.

An alternative explanation for the transition in $\alpha_\mathrm{B}$ could lie in the fact that plasma at greater radial distances has had a greater number of nonlinear times pass during its journey from the Sun. It, therefore, might be argued that the transition is reflective of the turbulence evolving during its journey from an earlier transient state. \cite{2021A&A...650A..21S} suggested that the turbulence age, a parameter characterising this effect, could be behind the transition after finding variation of the index with both solar wind speed and radial distance. However, \cite{2020ApJS..246...53C} found that, for distances as close as $35.7 \, R_{\odot}$, the travel time from the Sun is much greater than the outer scale nonlinear time so the turbulence should already be well evolved.

As the parameters discussed above themselves vary with distance it is possible that $\alpha_\mathrm{B}$'s apparent dependence on those parameters is merely a reflection of the parameters' and $\alpha_\mathrm{B}$'s shared dependence on distance. In this paper a statistical analysis is presented, which, for the first time, rigorously separates the dependence of $\alpha_\mathrm{B}$ on distance from $\alpha_\mathrm{B}$'s dependence on other properties of the solar wind, in order to clearly identify which is controlling its behaviour and therefore the nature of the MHD inertial range in the solar wind.

\section{Data}
PSP data from orbits 1 to 11 were used, covering the date range 1st October 2018 to 31st March 2022. The magnetic field data were provided by the fluxgate magnetometer (MAG) of the FIELDS instrument suite \citep{2016SSRv..204...49B}, with the 4 samples per cycle data product being used throughout this paper. The ion velocity data were provided by the SPAN-I instrument of the SWEAP suite \citep{2016SSRv..204..131K}, with bi-Maxwellian fits \citep{2021A&A...650L...1W} used during encounters 2 to 7, where available, and moments being used otherwise. Fits data were only used where at least 3 $\phi$ bins were fitted to, in order to ensure that the proton core was sufficiently captured. Density data were obtained from the quasi-thermal noise (QTN) measurements made by the Radio Frequency Spectrometer Low Frequency Receiver \citep{2020ApJS..246...44M}. Density data from SPAN-I were also used but only as a check on the quality of the velocity data, as described in Section 3.2.

\newpage

\section{Results}
\subsection{Dependence of magnetic spectral index with distance}
The magnetic field data were divided into intervals of six hour duration in order to study the dependence of the spectral index, $\alpha_\mathrm{B}$, on heliocentric distance, $r$. Only intervals where PSP was at a heliocentric distance of less than $150 \, R_{\odot}$ were considered and any intervals with more than 1\% of data points missing were excluded from the analysis. This left 1873 intervals. For each interval a fast Fourier transform was performed to produce a trace power spectral density. Invoking the Taylor hypothesis \citep{1938RSPSA.164..476T} allows such frequency spectra to be interpreted as wavenumber spectra. \cite{2021A&A...650A..22P} found the Taylor hypothesis to be appropriate for analysis with PSP data, even when working with data from its closest approaches to the Sun. $\alpha_\mathrm{B}$ was calculated for each interval in the spacecraft-frame frequency range $10^{-2}\,\mathrm{Hz} < f_\mathrm{sc} < 10^{-1}\, \mathrm{Hz}$, it was verified that this corresponds to the MHD inertial range for each interval used. All the analysis in this paper involving the magnetic spectral index was repeated with $\alpha_\mathrm{B}$ calculated over a range of a fixed number of ion gyroradii (assuming the Taylor hypothesis), this was found to have no significant impact on the results.      

\begin{figure}[ht!]
\centering
\includegraphics[width=\columnwidth,trim=0 0 0 0,clip]{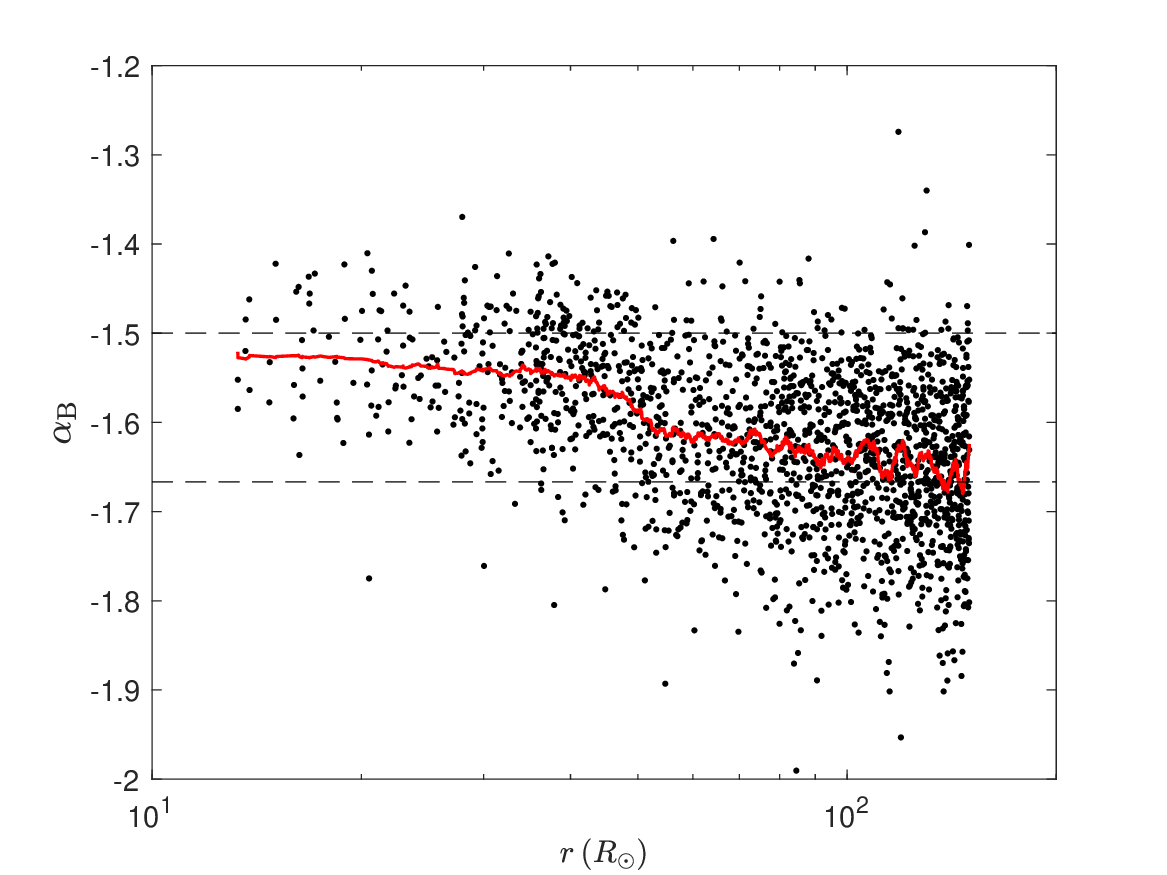}
\caption{The trace magnetic field spectral index, $\alpha_\mathrm{B}$, of six hour PSP intervals against the heliocentric distance, $r$, with $\alpha_\mathrm{B}$ calculated in the frequency range $10^{-2}\,\mathrm{Hz} < f_\mathrm{sc} < 10^{-1}\, \mathrm{Hz}$. The red line is a 75-point running mean. The dashed lines mark the spectral index values commonly predicted from theory, $-3/2$ and $-5/3$.} 
\label{fig:6hr_index}
\end{figure}

The index, $\alpha_\mathrm{B}$, calculated for each interval is shown in Figure \ref{fig:6hr_index} as a function of heliocentric distance, $r$. At large distances the results are consistent with a $-5/3$ scaling but are close to a $-3/2$ scaling at the closest distances to the Sun. The transition between the two values occurs at about $50 \, R_{\odot}$. This result is in agreement with \cite{2020ApJS..246...53C}, with the additional finding that the index appears to saturate near $-3/2$ as the Sun is approached.

\begin{figure}[ht!]
\centering
\includegraphics[width=\columnwidth,trim=0 0 0 0,clip]{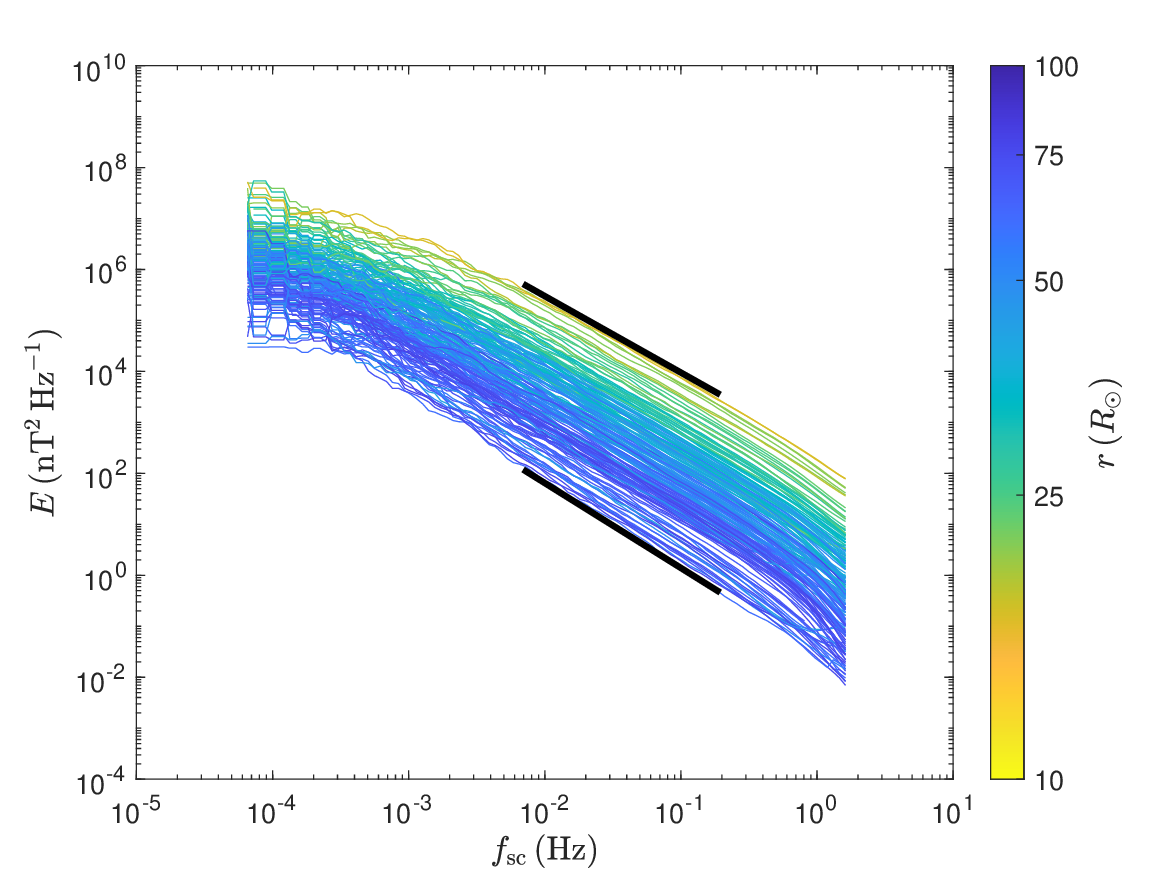}
\caption{Example smoothed, six hour magnetic field power spectra, $E$, coloured by the heliocentric distance, $r$, of the intervals the spectra are calculated from. The solid black lines mark slopes corresponding to spectral index values of $-3/2$ and $-5/3$.}
\label{fig:spectra_r}
\end{figure}

The transition is further illustrated in Figure \ref{fig:spectra_r}. A selection of trace power spectra from the intervals are shown, with the colour of the spectra indicating the heliocentric distance at which they were measured. The spectra have been smoothed by averaging over a sliding window of a factor of two. Consistent with the above discussion, the spectra measured closest to the Sun are clearly shallower than those at the greatest distances and are consistent, in their inertial range, with a $-3/2$ scaling indicated by the upper solid black line. The spectra measured at the furthest distances are consistent with a $-5/3$ scaling, indicated with the lower solid black line.

\subsection{Dependence on cross helicity and residual energy}
To investigate the mechanism behind the transition in the value of $\alpha_\mathrm{B}$, other parameters of the solar wind, plausibly underlying the transition, were also measured. In order to determine which, if any, of these parameters may be responsible for the transition, the variation of $\alpha_\mathrm{B}$ with distance, $r$, was separated from its variation with these parameters. 

Those considered include the normalised cross helicity, defined as

\begin{equation} \label{eq:cross_helicity}
\sigma_{\mathrm{c}} = \frac{\langle \delta \boldsymbol{z}^{+2} - \delta \boldsymbol{z}^{-2} \rangle}{\langle \delta \boldsymbol{z}^{+2} + \delta \boldsymbol{z}^{-2} \rangle},
\end{equation}
and the normalised residual energy, defined as

\begin{equation} \label{eq:res_energy}
\sigma_{\mathrm{r}} = \frac{2 \langle \delta \boldsymbol{z}^{+2} \cdot \delta \boldsymbol{z}^{-2} \rangle}{\langle \delta \boldsymbol{z}^{+2} + \delta \boldsymbol{z}^{-2} \rangle}, 
\end{equation}
where the angular brackets represent averages taken over the interval.
The imbalance and alignment of the Elsasser fields, characterised by $\sigma_{\mathrm{c}}$ and $\sigma_{\mathrm{r}}$ respectively, are a factor in determining the magnitude of the non-linear term in Equation (\ref{eq:elasser_evol}), the governing equation of ideal MHD turbulence. This, and their known radial dependence \citep{1987JGR....9211021R, 1995SSRv...73....1T, 1998JGR...103.6521B, 2000JGR...10515959B, 2004GeoRL..3112803M, 2005GeoRL..32.6103B, 2007AnGeo..25.1913B, 2020ApJS..246...53C}, make $\sigma_{\mathrm{c}}$ and $\sigma_{\mathrm{r}}$ clear candidates for a potential parameter underlying the $\alpha_\mathrm{B}$ transition. 

The data were divided into one hour intervals. Only intervals where PSP was at heliocentric distance of less than $80 \, R_{\odot}$ were considered. Any interval with at least 1\% of the magnetic field data, 10\% of the ion velocity data or 80\% of the density data missing was discarded. Intervals where the average SPAN-I measured density was less than 10\% of the density measured from quasi-thermal noise were also discarded. This final condition, along with the condition on the heliocentric distances of the intervals, is to ensure the SPAN-I measurements are sufficiently capturing the velocity distribution of the solar wind, which is not always fully in the instrument's field of view \citep{2016SSRv..204..131K}. After application of these conditions, 1894 intervals remained, of which 558 obtained their velocity data from bi-Maxwellian fits, the remainder from moments.

\begin{figure*}[ht!]
\centering
\includegraphics[width=0.9\textwidth,trim=0 0 0 0,clip]{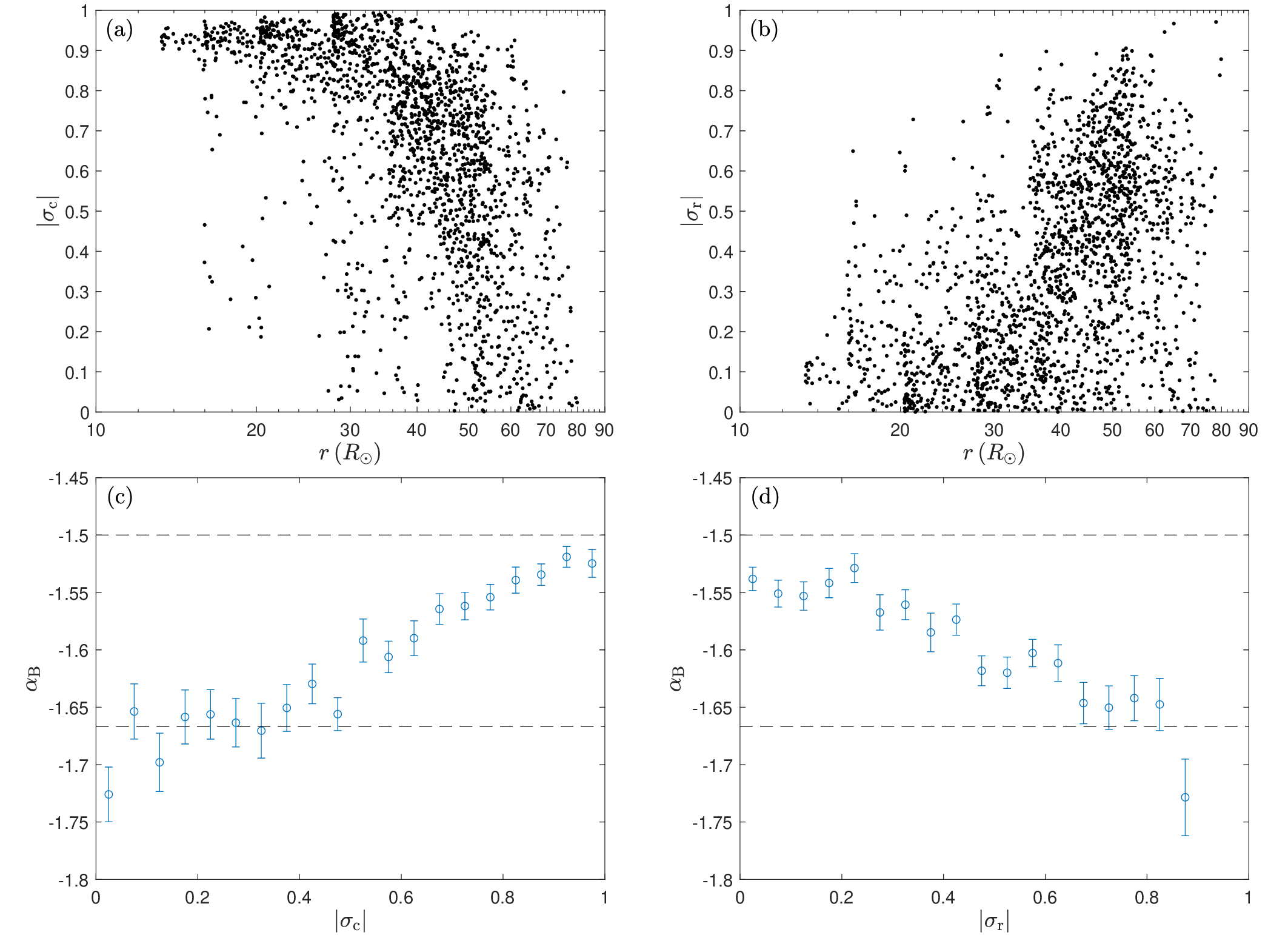}
\caption{(a) The absolute cross helicity, $|\sigma_{\mathrm{c}}|$, dependence and (b) the absolute residual energy,  $|\sigma_{\mathrm{r}}|$, dependence on heliocentric distance, $r$. (c) The mean magnetic spectral index, $\alpha_\mathrm{B}$, for intervals binned by $|\sigma_{\mathrm{c}}|$ with associated standard errors. The dashed lines mark $\alpha_\mathrm{B}$ values of $-3/2$ and $-5/3$. (d) The equivalent for $|\sigma_{\mathrm{r}}|$ bins.}
\label{fig:sigmas}
\end{figure*}

Both $\sigma_{\mathrm{c}}$ and $\sigma_{\mathrm{r}}$ were calculated in the inertial range. This was achieved by determining the perturbations to the Elsasser variables in Equations \ref{eq:cross_helicity} and \ref{eq:res_energy} using  
$\delta \boldsymbol{z}^{\pm}(t) = \boldsymbol{z}^{\pm}(t+\tau) - \boldsymbol{z}^{\pm}(t)$, with $\tau \approx100\,\mathrm{s}$, a duration that corresponds to the inertial range. $\boldsymbol{z}^{\pm}(t)$ were calculated using only the magnetic field and ion velocity components perpendicular to $\boldsymbol{B}_0$, the mean magnetic field of each interval. Figure \ref{fig:sigmas}(a) and (b) show $|\sigma_{\mathrm{c}}|$ and $|\sigma_{\mathrm{r}}|$ as functions of $r$. The radial dependence of both quantities is immediately apparent with intervals of high imbalance and low residual energy being more frequent closer to the Sun. Note that when calculated with moments $|\sigma_{\mathrm{c}}|$ tended to be slightly lower than when calculated with the bi-Maxwellian fits. The apparent decrease in $|\sigma_{\mathrm{c}}|$ as the Sun is approached at the smallest $r$ displayed, where only moments are available, is therefore possibly artificial.

For each interval $\alpha_\mathrm{B}$ was calculated, as described in the previous section. The intervals were binned by absolute cross helicity or absolute residual energy and the mean index for each bin determined, with associated standard error. The results are shown in \ref{fig:sigmas}(c) and (d) for $|\sigma_{\mathrm{c}}|$ and $|\sigma_{\mathrm{r}}|$, respectively. The clear trends of increasing index for increasing imbalance and decreasing index for increasing absolute residual energy are consistent with the trends of $\alpha_\mathrm{B}$, $|\sigma_{\mathrm{c}}|$ and $|\sigma_{\mathrm{r}}|$ with $r$. These strong trends make both $|\sigma_{\mathrm{c}}|$ and $|\sigma_{\mathrm{r}}|$ good candidates for the analysis of this paper.

\begin{figure*}[ht!]
\centering
\includegraphics[width=0.9\textwidth,trim=0 0 0 0,clip]{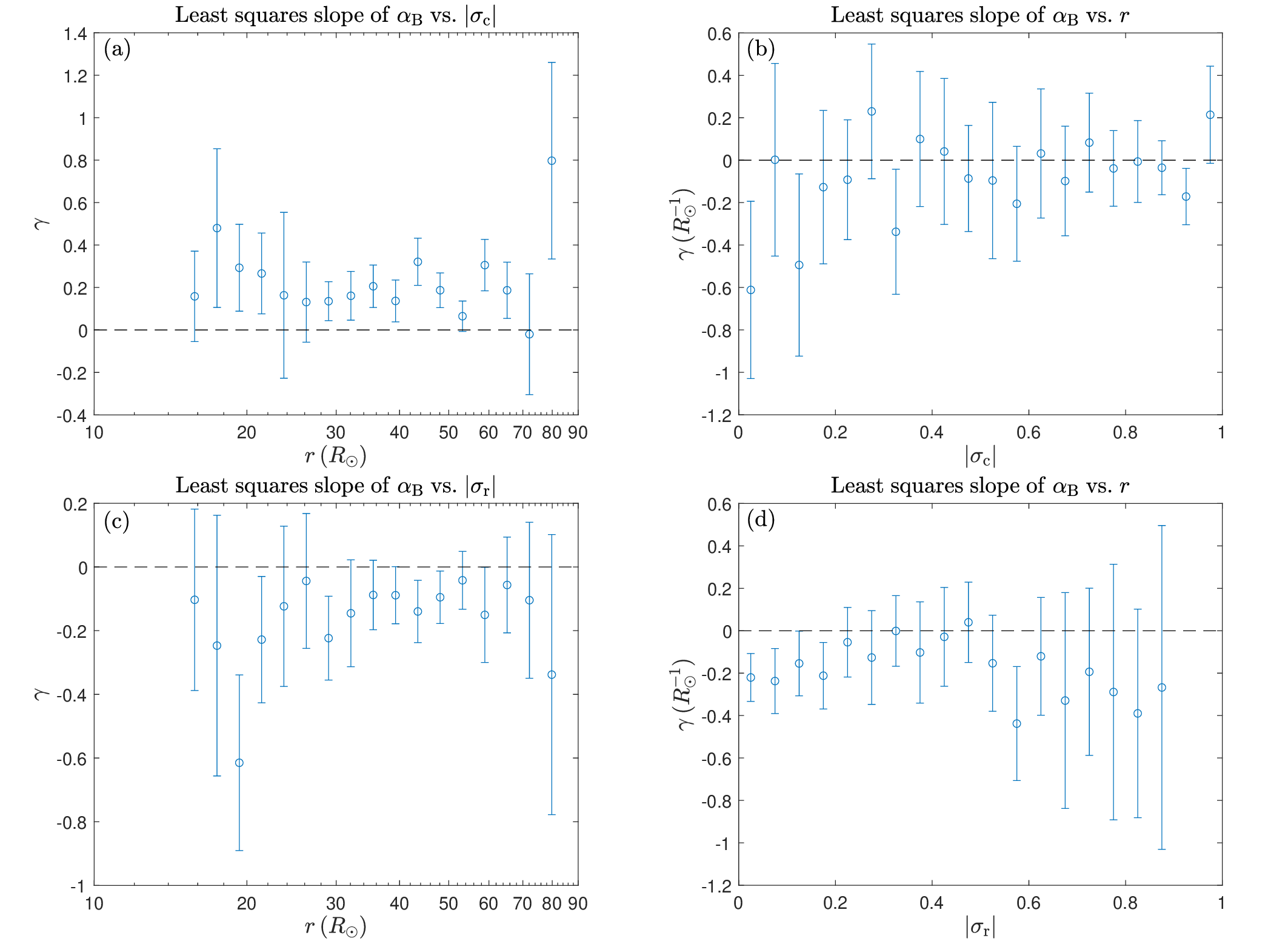}
\caption{(a) The gradient from a least squares linear fit, $\gamma$, of the magnetic spectral index, $\alpha_{\mathrm{B}}$, against absolute cross helicity, $|\sigma_c|$, for sets of intervals binned by heliocentric distance, $r$. (b) $\gamma$ of $\alpha_\mathrm{B}$ against $r$ for sets of intervals binned by $|\sigma_c|$. (c) $\gamma$ of $\alpha_\mathrm{B}$ against absolute residual energy, $|\sigma_r|$, for sets of intervals binned by $r$. (d) $\gamma$ of $\alpha_\mathrm{B}$ against  $r$ for sets of intervals binned by $|\sigma_r|$. 95\% confidence intervals for each $\gamma$ are marked. In each case a dotted line marks $\gamma=0$ to illustrate the statistical significance, or lack thereof, of each $\gamma$. The data points for bins containing fewer than 10 intervals are not shown.}

\label{fig:sigmas_error}
\end{figure*}

To examine whether $\sigma_{\mathrm{c}}$, say, may be behind the transition in $\alpha_\mathrm{B}$, the variation of $\alpha_\mathrm{B}$ with $r$ was separated from the variation of $\alpha_\mathrm{B}$ with $|\sigma_{\mathrm{c}}|$. In order to do this one of $r$ or $|\sigma_{\mathrm{c}}|$ was held approximately constant and the response of $\alpha_\mathrm{B}$ to varying the other under this constraint was observed. Take first isolating the variation of $\alpha_\mathrm{B}$ with $|\sigma_{\mathrm{c}}|$ from its variation with $r$. The intervals were binned according to $r$ and, within each bin, a linear fit was performed of $\alpha_\mathrm{B}$ against $|\sigma_{\mathrm{c}}|$. For each bin the gradient of the linear fit, $\gamma$, and associated 95\% confidence interval from that fit are displayed in Figure \ref{fig:sigmas_error}(a), against the arithmetic centre of the heliocentric distance range of that bin. 12 of the confidence intervals do not contain zero and so have an associated $\gamma$ statistically different from zero. This is strong evidence that, even when $r$ is kept approximately constant, $\alpha_\mathrm{B}$ continues to vary with $\sigma_{\mathrm{c}}$.  

To isolate the variation with  $r$ from the variation with $\sigma_{\mathrm{c}}$ a similar procedure was followed. In this case the intervals were binned by $|\sigma_{\mathrm{c}}|$ and, within each bin, a linear fit of $\alpha_{\mathrm{B}}$ to  $r$ was performed. Again a gradient, $\gamma$, and associated 95\% confidence interval were obtained, the results are shown in \ref{fig:sigmas_error}(b). In this case only 4 of the bins have an associated $\gamma$ which is statistically different from zero and the sign of $\gamma$ is inconsistent across bins. There is therefore little evidence of a trend with $r$ remaining when $|\sigma_{\mathrm{c}}|$ is held approximately constant. Figures \ref{fig:sigmas_error}(a) and (b) therefore suggest cross helicity is a strong candidate for a parameter underlying the observed transition in $\alpha_\mathrm{B}$.

The above process was repeated with $|\sigma_{\mathrm{r}}|$ and $r$, the results are shown in Figures \ref{fig:sigmas_error}(c) and (d). The data points for bins containing fewer than 10 intervals are not shown and are excluded from analysis, as is the case throughout this paper. Figure \ref{fig:sigmas_error}(c) is analogous to \ref{fig:sigmas_error}(a) with the intervals binned by $r$ to isolate the variation of $\alpha_{\mathrm{B}}$ with $|\sigma_{\mathrm{r}}|$. 6 of the bins have confidence intervals that do not contain zero. There is, therefore, weaker evidence that the trend of $\alpha_{\mathrm{B}}$ with $|\sigma_{\mathrm{r}}|$ remains when $r$ is held approximately constant compared to the $|\sigma_{\mathrm{c}}|$ case. Figure \ref{fig:sigmas_error}(d) is analogous to \ref{fig:sigmas_error}(b); the intervals are binned by $|\sigma_{\mathrm{r}}|$  to isolate the effect of varying $r$ on the index. Only 5 of the bins have associated $\gamma$ with confidence intervals that do not include zero and therefore there is some evidence that holding $|\sigma_{\mathrm{r}}|$ constant has removed the apparent trend with $r$. Overall, Figures \ref{fig:sigmas_error}(c) and (d) suggest some evidence in favour of residual energy as a candidate for a parameter underlying the observed transition but this evidence is weaker than that for the cross helicity.

\begin{figure}[ht!]
\centering
\includegraphics[width=\columnwidth,trim=0 0 0 0,clip]{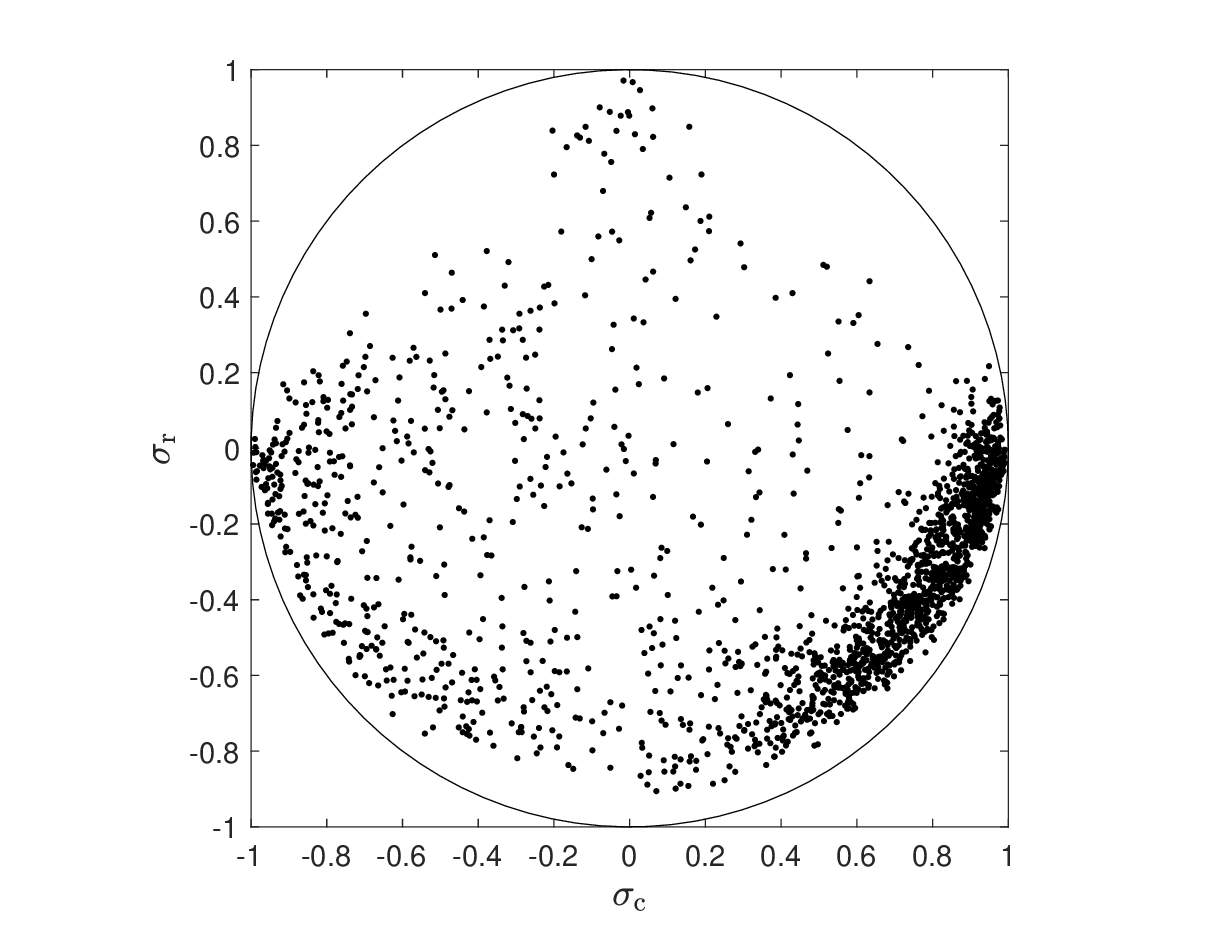}
\caption{The dependence of the residual energy, $\sigma_{r}$, on the cross helicity, $\sigma_{c}$. The circle marks $\sigma_{c}^2 + \sigma_{r}^2 = 1$. Note that a positive $\sigma_{\mathrm{c}}$ does not always correspond to an excess of energy in the Elsasser variable propagating away from the Sun, as the direction of  $\boldsymbol{B}$ has not been altered in the calculation of $\sigma_{c}$ to ensure this.}
\label{fig:sigmas_circle}
\end{figure}

\begin{figure*}[p!]
\centering
\includegraphics[width=0.9\textwidth,trim=0 0 0 0,clip]{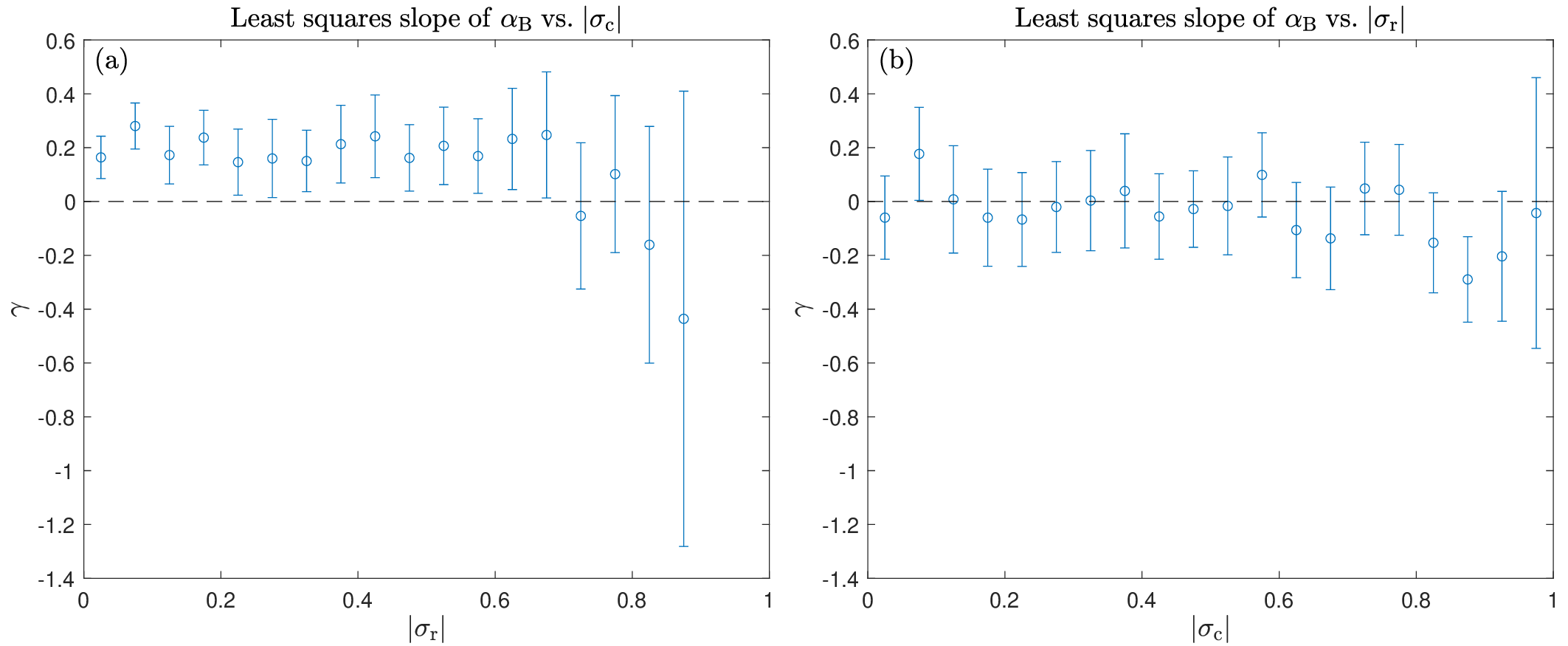}
\caption{(a) The gradient from a least squares linear fit, $\gamma$, of the spectral index, $\alpha_\mathrm{B}$, against absolute cross helicity, $|\sigma_c|$, for sets of intervals binned by absolute residual energy, $|\sigma_r|$. (b) $\gamma$ of $\alpha_\mathrm{B}$ against $|\sigma_r|$ for sets of intervals binned by $|\sigma_c|$. 95\% confidence intervals for each $\gamma$ are marked. The data points for bins containing fewer than 10 intervals are not shown.} 
\label{fig:circle_error}
\end{figure*}

It should be noted that $\sigma_{\mathrm{c}}$ and $\sigma_{\mathrm{r}}$ do not vary independently and so it is possible that an apparent trend with one is due to the trend with the other. The values of $\sigma_{\mathrm{r}}$ and $\sigma_{\mathrm{c}}$ for the intervals used are plotted against each other in Figure \ref{fig:sigmas_circle}.  From Equations (\ref{eq:cross_helicity}) and (\ref{eq:res_energy}) it is apparent that $\sigma_{\mathrm{c}}^2 + \sigma_{\mathrm{r}}^2 \le 1$. There is a tendency, observed in previous studies \citep{2006AnGeo..24.3179B, 2007AnGeo..25.1913B, 2010ApJ...717..474D, 2013ApJ...770..125C, 2013ApJ...778..177W}, for the points to preferentially lie towards the edge of the circle this condition defines and cluster in the negative residual energy, positive cross helicity quadrant. Given this, it is important to separate the dependence of $\alpha_{\mathrm{B}}$ on $\sigma_{\mathrm{c}}$ and on $\sigma_{\mathrm{r}}$. The above analysis technique was therefore used with $|\sigma_{\mathrm{c}}|$ and $|\sigma_{\mathrm{r}}|$, $r$ no longer being considered. The results are shown in Figure \ref{fig:circle_error}. Figure \ref{fig:circle_error}(a) shows $\gamma$ for a fit of $\alpha_\mathrm{B}$ against $|\sigma_{\mathrm{c}}|$ for intervals binned by $|\sigma_{\mathrm{r}}|$. For 14 of the bins the confidence interval does not include zero. This indicates that, even with $|\sigma_{\mathrm{r}}|$ held constant, there is still good evidence of statistically significant variation of $\alpha_\mathrm{B}$ with $|\sigma_{\mathrm{c}}|$. Figure  \ref{fig:circle_error}(b) shows the reverse, with $\gamma$ corresponding to a fit of $\alpha_\mathrm{B}$ against $|\sigma_{\mathrm{r}}|$ for intervals binned by $|\sigma_{\mathrm{c}}|$. In this case only two of the intervals have a corresponding $\gamma$ with a confidence interval that does not include zero. When $|\sigma_{\mathrm{c}}|$ is held constant it appears the trend with $|\sigma_{\mathrm{r}}|$ vanishes. This suggests that the apparent trend with residual energy is simply a manifestation of the underlying trend with cross helicity.

\subsection{Dependence on turbulence age}
A parameter also considered was the turbulence age  \citep{1998JGR...103.6495M}, the approximate number of outer scale nonlinear times that have passed for a parcel of plasma during its journey from the Sun, as it is possible that the trend observed in $\alpha_\mathrm{B}$ may be due to the turbulence evolving in time as it becomes fully developed.

\begin{figure*}[ht!]
\centering
\includegraphics[width=0.9\textwidth,trim=0 0 0 0,clip]{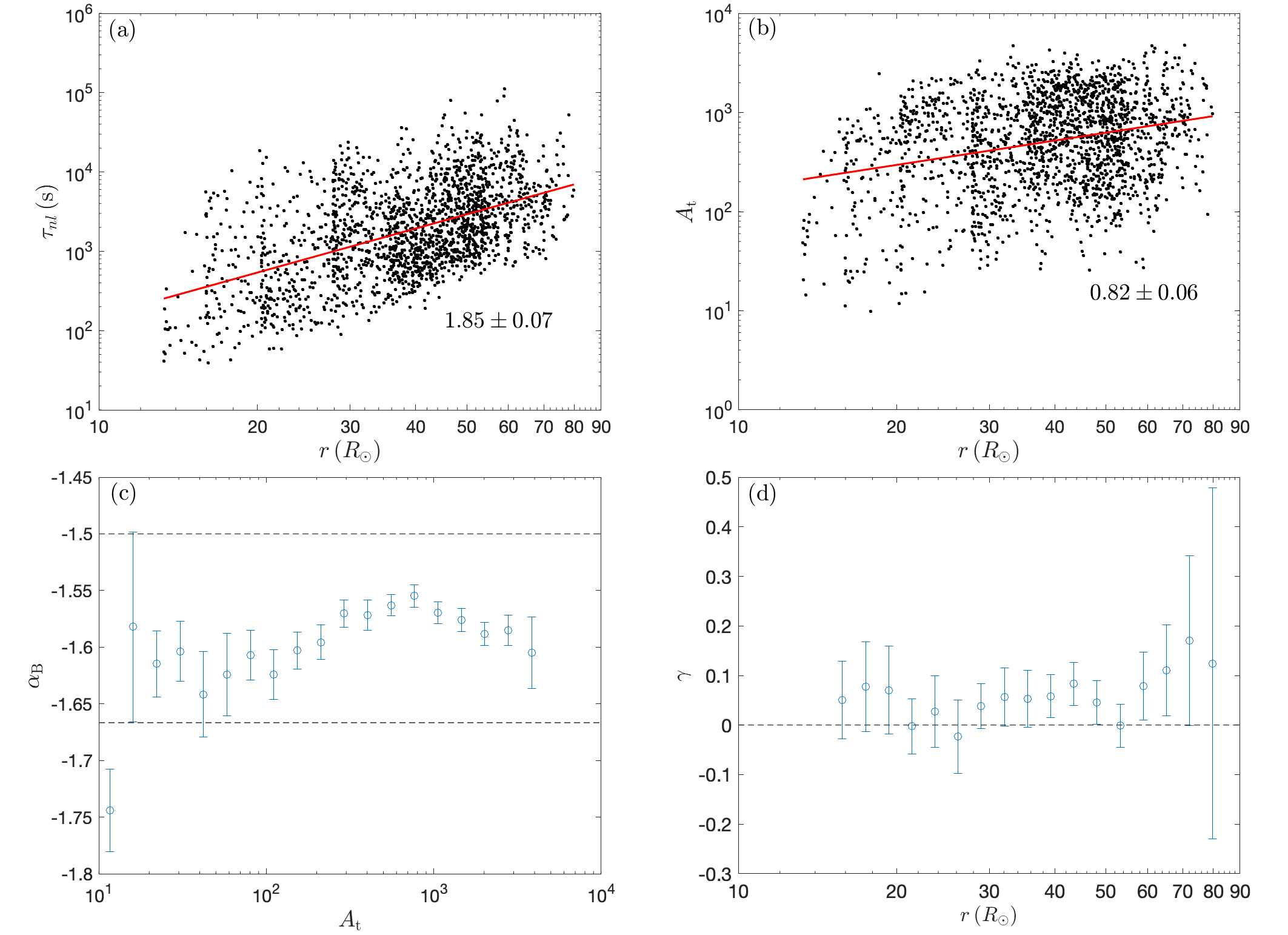}
\caption{(a) The dependence of the nonlinear time, $\tau_{\mathrm{nl}}$, on heliocentric distance, $r$. (b) The dependence of the turbulence age, $A_\mathrm{t}$, on $r$ with the age calculated using Equation \ref{eq:turb_age3}. (c) The mean spectral index, $\alpha_\mathrm{B}$, for intervals binned by $A_\mathrm{t}$ with associated standard errors. (d) The gradient from a least squares linear fit, $\gamma$, of $\alpha_\mathrm{B}$ against $A_\mathrm{t}$ for sets of intervals binned by $r$ with 95\% confidence intervals.}
\label{fig:turb_age}
\end{figure*}

Equation (\ref{eq:elasser_evol}) suggests a form for the nonlinear time of $\tau_{\mathrm{nl}} \sim {\lambda}/{\delta b}$, where $\lambda$ is the scale of the fluctuation and $\delta b$ is in velocity units. Note that this form of $\tau_{\mathrm{nl}}$ does not account for effects arising from alignment or imbalance of the Elsasser fields. For this paper $\lambda$ was taken to be the correlation scale, measured as the time scale over which the correlation function, 
\begin{equation} \label{eq:corr_funct}
C(\tau)= \langle \delta \boldsymbol{B}(t+\tau) \cdot \delta \boldsymbol{B}(t) \rangle,
\end{equation}
where $\delta \boldsymbol{B}(t) = \boldsymbol{B}(t) - \langle \boldsymbol{B} \rangle$, decreases by a factor of $e$; which was then converted to a length scale using the Taylor hypothesis \citep{2015JGRA..120..868I,2020ApJS..246...53C}. $\delta b$ was taken to be the square root of the value of the magnetic field second-order structure function,
\begin{equation} \label{eq:struct_funct}
S_{2}(\tau) = \langle |\boldsymbol{B}(t+\tau) - \boldsymbol{B}(t)|^2 \rangle,
\end{equation}
at large scales, where it reaches a steady value, in velocity units \citep{2020ApJS..246...53C}. 
The resulting $\tau_{\mathrm{nl}}$ for each of the 1894 intervals used in the previous section is shown in Figure \ref{fig:turb_age}(a). As the correlation scale increases with distance and the magnetic fluctuation amplitudes decrease, the outer scale nonlinear time is seen to increase with distance from the Sun.

If $\tau_{\mathrm{nl}}$ is taken to be constant for a given plasma parcel over its journey from the Sun, then the turbulence age is estimated as $A_\mathrm{t} = T/\tau_{\mathrm{nl}}$, where $T$ is the travel time from the Sun. However, calculated this way, $\tau_{\mathrm{nl}}$ was found to increase at such a rate with distance that $A_\mathrm{t}$ would decrease with distance, which clearly cannot  be correct. The assumption that the nonlinear time is constant was therefore abandoned and the following integral instead considered,

\begin{equation} \label{eq:turb_age1} 
A_\mathrm{t}(t) =  \int_{0}^{t} \frac{dt'}{\tau_{\mathrm{nl}}(t')}. 
\end{equation}
Taking the solar wind speed, $V_{\mathrm{sw}}$, to be constant with distance and performing a change of variables gives,

\begin{equation} \label{eq:turb_age2}
A_\mathrm{t}(r) =  \frac{1}{V_{\mathrm{sw}}} \int_{r_0}^{r} \frac{dr'}{\tau_{\mathrm{nl}}(r')}. 
\end{equation}

It was then assumed that $\tau_{\mathrm{nl}}$ follows a power law, $\tau_{\mathrm{nl}} \propto r^{a}$. Figure \ref{fig:turb_age}(a) gives justification to this assumption, with $\tau_{\mathrm{nl}}$ appearing reasonably well captured by a such a function. If $\tau_{\mathrm{nl}}$ is measured at some distance $r_\mathrm{m}$ to be $\tau_{\mathrm{nl,m}}=\tau_{\mathrm{nl}}(r=r_\mathrm{m})$ then $\tau_{\mathrm{nl}} = (r/r_\mathrm{m})^{a}\tau_{\mathrm{nl,m}}$. Performing the integral gives,

\begin{equation} \label{eq:turb_age3}
A_\mathrm{t}(r) =  \frac{1}{1-a} \frac{r_\mathrm{m}^{a}}{V_{\mathrm{sw}}\tau_{\mathrm{nl,m}}}[r^{1-a}-r_0^{1-a}]. 
\end{equation}

The value of $a$ was calculated to be 1.85 by performing a fit of $\tau_{\mathrm{nl}}$ against distance as shown in Figure \ref{fig:turb_age}(a). This value was used for all intervals. For each interval $r_\mathrm{m}$ and $\tau_{\mathrm{nl,m}}$ were taken to be the values as calculated for that interval, as was the case for $V_{\mathrm{sw}}$. A value had to be set for $r_0$, $13R_{\odot}$ was used for each interval --- this being a value below all heliocentric distances of the intervals used. While setting a value too far from the Sun will result in a systematically underestimated  $A_\mathrm{t}$, note that what is important for this present analysis is the relative  $A_\mathrm{t}$ between points, rather than the absolute  $A_\mathrm{t}$. The results are shown in  \ref{fig:turb_age}(b), with a least squares fit demonstrating that  $A_\mathrm{t}$ increases with distance. $A_\mathrm{t} \gg 1$ for all intervals, consistent with \cite{2020ApJS..246...53C}, which would suggest well developed turbulence, and so appears to undermine the suggestion that the turbulence age may be behind the transition, though, as stated above, the form of $\tau_{\mathrm{nl,m}}$ used here does not take into account imbalance or alignment of the Elsasser fields.

The intervals were binned by the calculated $A_\mathrm{t}$ and the mean $\alpha_\mathrm{B}$ for each bin determined with associated standard error, the results are shown in Figure \ref{fig:turb_age}(c). Unlike in the cases of the trend with $r$, $\sigma_{\mathrm{c}}$ or $\sigma_{\mathrm{r}}$, there is no clear trend of $\alpha_\mathrm{B}$ with $A_\mathrm{t}$. Nevertheless, the analysis of the previous section was repeated to attempt to separate any dependence of $\alpha_\mathrm{B}$ on $A_\mathrm{t}$ from the dependence on $r$. Analogous to the above analysis, the intervals were binned according to distance and $\gamma$ was calculated for each, with associated 95\% confidence intervals, and is shown in Figure \ref{fig:turb_age}(d). Only 5 bins have associated error bars do not include zero. From this, and Figure \ref{fig:turb_age}(c), it follows that the evidence for turbulence age being the parameter underlying the transition in the spectral index is far weaker than is the case for cross helicity. 

\subsection{Dependence on further parameters}
Other parameters possibly underlying the variation in $\alpha_{\mathrm{B}}$ were considered and the above statistical analysis repeated for each. The 1894 intervals of the previous two sections were used for each parameter examined.

\begin{figure*}[ht!]
\centering
\includegraphics[width=\textwidth,trim=0 0 0 0,clip]{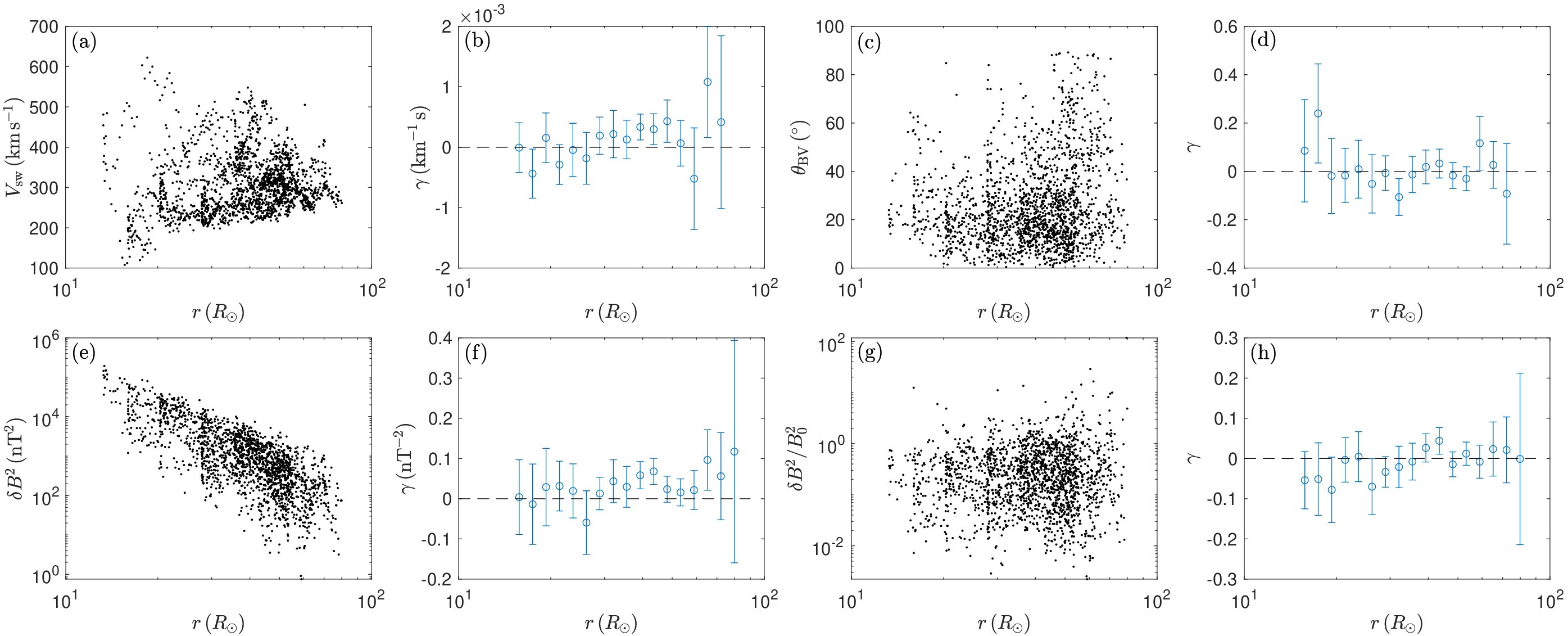}
\caption{The dependence of (a) the solar wind speed, $V_{\mathrm{sw}}$; (c) the sampling angle, $\theta_{\mathrm{BV}}$; (e) the magnitude of the magnetic field fluctuations, $\delta B$, squared and (g) the normalised fluctuation magnitude, $\delta B/B_0$, squared on the heliocentric distance, $r$. (b) The gradient from a least squares linear fit, $\gamma$, of the spectral index, $\alpha_\mathrm{B}$, against $V_{\mathrm{sw}}$ for sets of intervals binned by $r$, with 95\% confidence intervals, and the equivalent for $\gamma$ of $\alpha_\mathrm{B}$ against (d) $\theta_{\mathrm{BV}}$, (f) $\delta B^2$ and (h) $\delta B^2/B_0^2$. The data points for the largest $r$ bin in (b) and (d) are not shown due to having large associated confidence intervals, to allow the confidence intervals of other bins to be seen clearly.}
\label{fig:further_parameters}
\end{figure*}

\cite{2021A&A...650L...3C} found $\alpha_{\mathrm{B}}$ to depend on wind type, which the solar wind velocity is a common proxy for.
Further \cite{2021A&A...650A..21S} reported a trend of increasing index with increasing velocity. The mean solar wind velocity, $V_{\mathrm{sw}}$, against $r$ for each interval is shown in Figure \ref{fig:further_parameters}(a), clearly showing the acceleration of the solar wind from the Sun. Figure \ref{fig:further_parameters}(b) shows the results of separating any dependence of $\alpha_{\mathrm{B}}$ on $V_{\mathrm{sw}}$ from its apparent dependence on $r$, with the intervals being binned by $r$, and $\gamma$ with associated confidence interval being determined for each. 5 bins have associated confidence intervals that do not include zero and the sign of $\gamma$ is inconsistent across these bins. The evidence for $V_{\mathrm{sw}}$ as the underlying parameter is therefore weak.

A further parameter considered was the sampling angle --- the angle between the mean magnetic field and mean solar wind velocity in the spacecraft frame. Solar wind turbulence is known to be anisotropic \citep{2012SSRv..172..325H, chen_2016} meaning that different properties may be observed depending on the angle PSP's path makes with the background field, potentially explaining the observed index trend with distance. The measured angles, $\theta_{\mathrm{BV}}$, for each of the intervals used are shown against $r$ in Figure \ref{fig:further_parameters}(c). There is a clear trend with distance, with greater $\theta_{\mathrm{BV}}$ values tending to be observed further from the Sun. A similar analysis to the above is shown in Figure \ref{fig:further_parameters}(d), the intervals here again binned by $r$ to isolate any trend with $\theta_{\mathrm{BV}}$. Only 3 of the bins have $\gamma$ which are statistically different from zero and so there is little evidence for a trend with the sampling angle once the trend with distance is taken into account.

The magnitude of the magnetic field fluctuations, both unnormalised and normalised by the background field, were also considered. The latter is a factor in determining the turbulence strength and so may plausibly play a role in the transition of  $\alpha_\mathrm{B}$. $\delta B$ was calculated as in Section 3.3 but was not converted to velocity units. The unnormalised values against distance are shown in Figure \ref{fig:further_parameters}(e) and the normalised in Figure  \ref{fig:further_parameters}(g). While there is a very clear negative trend with distance in the unnormalised case there is no clear trend in the normalised case. Similar analysis to the above yields Figures \ref{fig:further_parameters}(f) and \ref{fig:further_parameters}(h) for the unnormalised and normalised case respectively. In the case of the unnormalised fluctuation magnitude only 3 bins have a $\gamma$ statistically different from zero, in the case of the normalised fluctuation it is only 2, and so the evidence for either underlying the transition in $\alpha_{\mathrm{B}}$ is weak. 

\subsection{The velocity field spectral index}
To aid with the interpretation of the above results, that point to cross helicity as the underlying parameter behind the dependence of the magnetic field spectral index with distance, the dependence of the velocity field spectral index, $\alpha_{\mathrm{v}}$, on cross helicity was considered. 

The lower cadence of the velocity measurements compared to the magnetic field measurements makes obtaining a good measure of $\alpha_{\mathrm{v}}$ considerably more difficult than obtaining a good measure of $\alpha_{\mathrm{B}}$. SPAN-I moments were used, rather than fits, to measure $\alpha_{\mathrm{v}}$ due to noise in the fits data at high frequencies. The data were divided into one hour intervals, only those with a resolution of at least 11 seconds were used. The selection criteria described in Section 3.2 were also applied. For each interval a fast Fourier transform was performed to produce a velocity power spectrum which was then smoothed by averaging over a sliding window of a factor of two. Many values for $\alpha_{\mathrm{v}}$ were then obtained by calculating $\alpha_{\mathrm{v}}$ over frequency ranges set by a sliding window, $0.2 \, f^{*} < f_\mathrm{sc} < f^{*}$, with $f^{*}$ ranging from $f_{\mathrm{max}}$, the maximum available frequency, down to $0.5 \, f_{\mathrm{max}}$. These ranges are selected as $f_{\mathrm{max}}$ is in the MHD inertial range for all intervals. The resulting set of indices were subject to a moving mean of a constant number of data points, with the variance associated with each mean recorded. The mean corresponding to the smallest variance was then selected as the final value for $\alpha_{\mathrm{v}}$ for the interval, the process being designed to select a frequency range to measure $\alpha_{\mathrm{v}}$ over which the value of $\alpha_{\mathrm{v}}$ is as close to constant as possible.

\begin{figure}[ht!]
\centering
\includegraphics[width=\columnwidth,trim=0 0 0 0,clip]{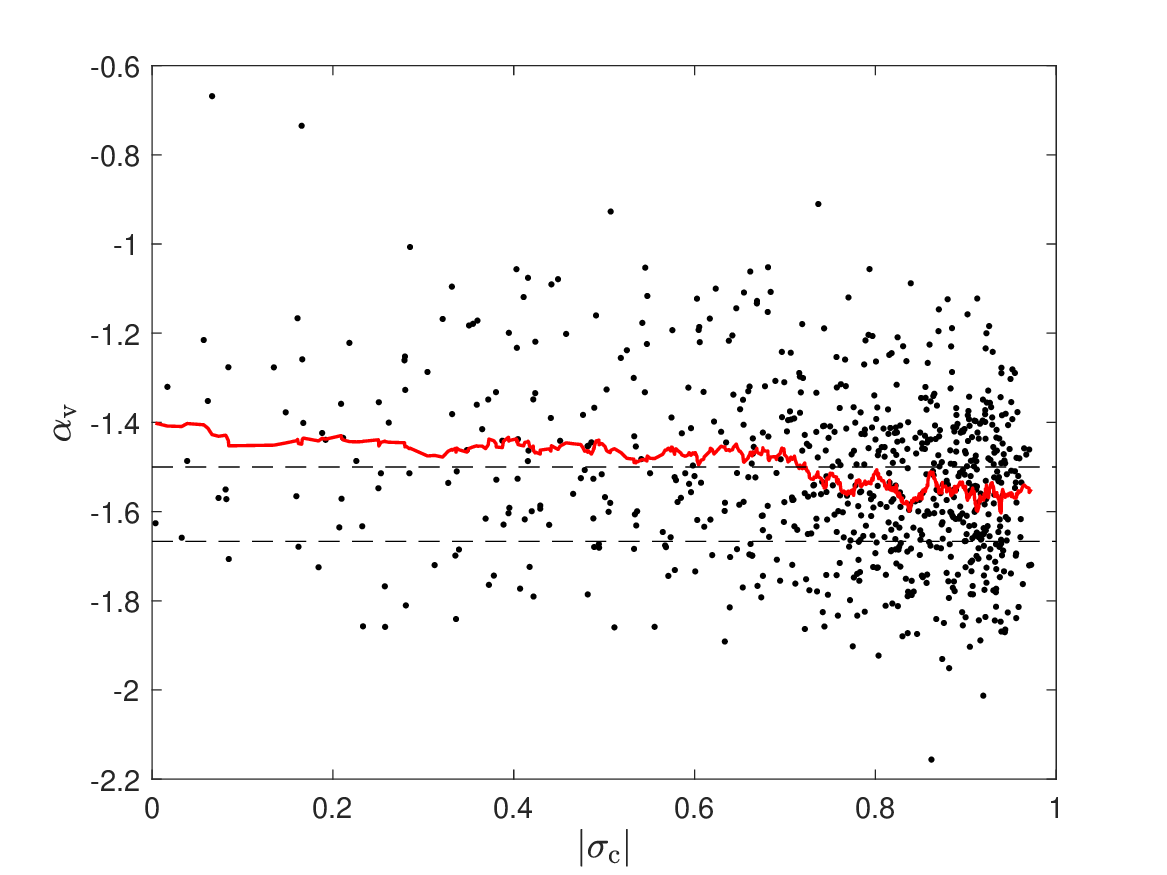}
\caption{The velocity field spectral index, $\alpha_{\mathrm{V}}$, of one hour PSP intervals against the absolute cross helicity, $|\sigma_{\mathrm{c}}|$. The red line is a 50-point running mean. The dashed lines mark the spectral index values $-3/2$ and $-5/3$.}
\label{fig:vel_index}
\end{figure}

The process was deemed to have performed sufficiently well when the minimum variance was below $10^{-4}$.  Discarding intervals where this was not the case left 757 intervals. An additional 12 intervals, for which unphysical values of $\alpha_{\mathrm{v}}$ were calculated and which contained heliospheric current sheet crossings or where the velocity distribution was not well captured, were also discarded. The measured $\alpha_{\mathrm{v}}$ against $|\sigma_{\mathrm{c}}|$ for the remaining intervals is shown in Figure \ref{fig:vel_index}, with $\sigma_{\mathrm{c}}$ determined as in previous sections. There is no evidence for a trend of the $\alpha_{\mathrm{v}}$ with $|\sigma_{\mathrm{c}}|$, with the running mean being consistent with $-3/2$ for all values of $|\sigma_{\mathrm{c}}|$. 

\section{Discussion}
In this paper a transition in the magnetic field spectral index has been shown from $-5/3$ far from the Sun to $-3/2$ close to the Sun, with the transition occurring at around $50R_{\odot}$. This is in agreement with previous observations  \citep{2020ApJS..246...53C, 2021A&A...650A..21S, 2023ApJ...943L...8S}. A saturation of $\alpha_{\mathrm{B}}$ at $-3/2$ as the Sun is approached is shown clearly for the first time. To gain insight into the physical mechanism responsible, the variation of the index with distance was separated from its variation from a number of other parameters plausibly responsible for the transition. Of all variables considered, the normalised cross helicity was found to be the only parameter to show a significant underlying effect on the spectral index. Previous work has found $\alpha_{\mathrm{B}}$ to vary with $\sigma_{\mathrm{c}}$ \citep{2010PhPl...17k2905P, 2013ApJ...770..125C, 2013ApJ...778..177W, 2018ApJ...865...45B, 2023ApJ...943L...8S}, this paper builds on those findings by rigorously isolating the variation with $\sigma_{\mathrm{c}}$ from variation with other parameters of the solar wind. This result contrasts with \cite{2018ApJ...865...45B}, who argued that the residual energy is the main controlling parameter, and \cite{2021A&A...650A..21S}, who argued for the turbulence age. However, the analysis presented here does not exclude the possibility of a secondary, weaker dependence on these parameters. There is no evidence for a similar trend for the velocity spectrum, which appears to be consistent with a $-3/2$ scaling regardless of the cross helicity. This is in agreement with observations at 1 au \citep{2010PhPl...17k2905P, 2013ApJ...770..125C, 2018ApJ...865...45B} and \cite{2021A&A...650A..21S}, which found no trend of $\alpha_{\mathrm{v}}$ with distance using PSP data. Some existing models of imbalanced turbulence do predict different behaviour for imbalanced compared to balanced turbulence \citep{2007ApJ...655..269L, 2008ApJ...685..646C, 2008ApJ...682.1070B, 2022JPlPh..88e1501S} but none predict the results obtained.

It is possible that excess magnetic energy in some regions, represented by a negative residual energy, could manifest as current sheets and \cite{2011PhRvL.106l5001L} found the presence of current sheets was associated with steeper magnetic spectra. This would be consistent with the found trend of the magnetic index on residual energy. In agreement with this potential connection \cite{2023arXiv230509763D} found discontinuities in the solar wind to be associated with steeper spectra. However, if such discontinuities were behind the transition in $\alpha_{\mathrm{B}}$ it would be expected that the dependence of $\alpha_{\mathrm{B}}$ on $|\sigma_{\mathrm{r}}|$ would be stronger than its dependence on $|\sigma_{\mathrm{c}}|$, which is the opposite to what has been found here. The apparent tendency for the residual energy to be maximised for a given cross helicity (Figure \ref{fig:sigmas_circle}) may provide a means by which the cross helicity could influence the index through this mechanism despite this.

An alternative explanation for the transition could lie in the potentially different behaviour of imbalanced, compared to balanced, turbulence. The imbalanced regions are, for example, where the theorised ``helicity barrier" is thought to be active \citep{2021JPlPh..87c5301M}. Under certain conditions a forward cascade of cross helicity meets a reverse cascade of magnetic helicity near the ion gyroscale, limiting the energy that can cascade forward for the dominant Elsasser field. The resulting buildup in energy at the gyroscale could result in a shallower spectrum, hence explaining the different observed scalings for different levels of imbalance. However, this would not account for why there is only a transition in the magnetic spectral index and not the velocity index, a challenge any potential explanation has to overcome.

The mechanism behind the observed behaviour of $\alpha_{\mathrm{B}}$ and $\alpha_{\mathrm{v}}$ remains an open question. The found strong dependence of the magnetic spectral index on the cross helicity perhaps points to an area where a new model of imbalanced MHD turbulence could be developed. Such a model would more fully capture the behaviour of the solar wind fluctuations than existing models and may better account for the impact of imbalance in MHD turbulence in general. 

\vspace{12pt}
JRM is supported by STFC studentship grant ST/V506989/1. CHKC is supported by UKRI Future Leaders Fellowship MR/W007657/1. CHKC and AL are supported by STFC Consolidated Grants ST/T00018X/1 and ST/X000974/1. JRM and AL acknowledge support from the Perren Exchange Programme. We thank Lloyd Woodham for providing the SPAN fits dataset and Alexander Schekochihin for helpful discussions. PSP data are available at the SPDF (https://spdf.gsfc.nasa.gov).

\bibliography{APJ_final}{}
\bibliographystyle{aasjournal}

\end{document}